\documentclass[noshowpacs,onecolumn,showpacs,superscriptaddress,groupedaddress,amsmath,amssymb]{elsarticle}
 \usepackage{graphicx}
 \usepackage{amsmath}
 \usepackage{amssymb}
 \usepackage{float}
 \usepackage{placeins}
 \usepackage{subfig}
 \usepackage{latexsym}
 \usepackage{bbold}
\usepackage[colorlinks]{hyperref}
\usepackage[left=2cm,right=2cm]{geometry}
 \usepackage{color}
 \usepackage{xcolor}
 \usepackage{float}
 \usepackage{makeidx}
 \usepackage{colortbl}
 \usepackage{csquotes}
\newcommand{\vect}[1]{\mathbf{#1}}

\biboptions{sort&compress}

 \title{Tunneling phase diagrams in anisotropic Multi-Weyl semimetals}

\begin{document}
\author[1]{Ahmed Bouhlal}
\author[2]{Adel Abbout}
\author[1]{Ahmed Jellal}
\author[2]{Hocine Bahlouli}
\author[2]{Michael Vogl}
\ead{ssss133@googlemail.com}
\address[1]{Laboratory of Theoretical Physics, Faculty of Sciences, Choua\"ib Doukkali University,  24000 El Jadida, Morocco}
\address[2]{Department of Physics, King Fahd University of Petroleum and Minerals, 31261 Dhahran, Saudi Arabia}
\begin{abstract}
Motivated by the exciting prediction of Multi-Weyl topological semimetals that are stabilized by point group symmetries \cite{PhysRevLett108266802}, we study tunneling phenomena for a class of anisotropic Multi-Weyl semimetals. We find that a distant detector for different ranges of an anisotropy parameter $\lambda$ and incident angle $\theta$ will measure a different number of propagating transmitted modes. We present these findings in terms of phase diagrams that is valid for an incoming wave with fixed wavenumber $k$--energy is not fixed. To gain a deeper understanding of this phenomenon we then focus on the simplest case of an anisotropic quadratic Weyl-semimetal and analyze tunneling coefficients analytically and numerically to confirm the observations from the phase diagram. Our results show non-analytical behavior, which is the hallmark of a phase transition. This serves as a motivation to make a formal analogy with phase transitions that are known from statistical mechanics. Specifically, we argue that the long distance limit in our tunneling problem takes the place of the thermodynamic limit in statistical mechanics. More precisely, find a direct formal connection to the recently developed formalism for dynamical phase transitions \cite{Heyl_2018}. We propose that this analogy to phase transitions can help classify transport properties in exotic semimetals.
\end{abstract}
\date{\today}
\maketitle
\tableofcontents
%%%%%%%%%%%%%%%%%%%%%%%%%%%%%%%%%%%%%%%%%%%%%%%%%%
\section{Introduction}
%%%%%%%%%%%%%%%%%%%%%%%%%%%%%%%%%%%
Recent theoretical and experimental advances have stimulated interest in classes of materials beyond the usual classification of metals, semiconductors, and insulators. Important newly discovered classes of materials include topological materials such as topological insulators and semimetals such as three-dimensional (3D) Dirac and Weyl semimetals, which are higher dimensional materials that share certain similarities with graphene \cite{Armitage_2018, doi:10.1146/annurev-conmatphys-031016-025458}. The class of topological materials possesses non-trivial protected topological properties such as edge states that are at the origin of their exciting electronic and transport properties \cite{hasan2010colloquium,qi2011topological,moore2010birth,tokura2019magnetic}. The characteristic feature of semimetals is that they are gapless but have neglibible overlap between conduction and valence band. That is, their conduction and valence bands touch at isolated points, nodal lines or nodal surfaces \cite{Fang_2016,xiao2020experimental,qie2019tetragonal,zhang2018nodal,xiao2017topologically,burkov2011topological,meng2020nonsymmorphic,wang2018pseudo,emmanouilidou2017magnetotransport,li2019new,huang2016topological,li2018evidence,yu2017topological,he2018type,li2018orthorhombic,laha2020magnetotransport,wu2018nodal,zhang2017topological,hosen2017tunability,chang2019realization}. Most exciting are cases, where these band crossings are topologically protected. Different types of topological semimetals can then be distinguished based on the degeneracy of the band crossings and the crystal space group symmetries that protects band crossings \cite{armitage2018weyl}. The qualitative behavior of the energy dispersion near the band crossings is also very important for transport behavior and can be considered as an additional discriminating characteristic of these materials. All together, these properties give rise to a wide class of distinct semimetals such as Dirac \cite{Armitage_2018, doi:10.1146/annurev-conmatphys-031016-025458, wang2017quantum,lundgren2014thermoelectric,lundgren2015electronic,hubener2017creating,wan2011topological,xu2011chern,fang2012multi,xiong2015evidence,liu2014stable,wang2013three,liang2015ultrahigh,novak2015large,wieder2016double,he2014quantum,jeon2014landau,weeks2010interaction,kim2015observation,li2018giant,chang2017type,szabo2020dirty,wieder2020strong,wang2012dirac,young2012dirac,liu2014discovery,yang2014classification,young2015dirac,borisenko2014experimental} or Weyl semimetals \cite{doi:10.1146/annurev-conmatphys-031016-025458, wang2017quantum, PhysRevX.7.021019,zyuzin2016intrinsic,jiang2017signature,wang2016mote,soluyanov2015type,huang2015weyl,xu2015discovery,weng2015weyl,huang2015observation,zyuzin2012weyl,lundgren2014thermoelectric,lundgren2015electronic,bulmash2014prediction,jiang2012tunable,burkov2011weyl,wang2020higher,ghorashi2020higher,wang2014floquet,zhang2016theory,hubener2017creating,chen2016thermoelectric,lv2015experimental,liu2019magnetic,vazifeh2013electromagnetic,Alidoust_2020,Alidoust_2018,Halterman_2019,Halterman_2018,Alidoust_2017,huang2016new,kane2005quantum,kane2005z,bernevig2006quantum,fu2007topological,haldane1988model,moore2007topological,bernevig2006quantum}, semimetals with higher order band crossings such as multi-Weyl- and Dirac semi-metals \cite{PhysRevX.7.021019,fang2012multi,huang2016new,chen2016thermoelectric,yang2014classification}, nodal line and surface semi-metals \cite{Fang_2016,xiao2020experimental,qie2019tetragonal,zhang2018nodal,xiao2017topologically,burkov2011topological,meng2020nonsymmorphic,wang2018pseudo,emmanouilidou2017magnetotransport,li2019new,huang2016topological,li2018evidence,yu2017topological,he2018type,li2018orthorhombic,laha2020magnetotransport,wu2018nodal,zhang2017topological,hosen2017tunability,chang2019realization}. 

However, there is still a large zoo of semimetals that have been predicted based on symmetry considerations but have not been exhaustively studied. In particular, we would like to point out to the recently predicted anisotropic cubic Dirac semi-metal \cite{liu2017predicted}. We have recently investigated theoretically a wealth of interesting tunnelling phenomena in cubic Dirac semimetals due to the richness of its band structure, which allows a linear energy dispersion in the $z$-direction and cubic energy dispersion and rotational symmetry, or lack of it in the $x$-$y$ plane. In general, Dirac or Weyl band crossings in bulk solids can be accompanied by different energy dispersions in different directions, as imposed by their crystal symmetries. At these band crossings, in general, the dispersion along the principle rotation axis is linear, whereas the dispersion along the perpendicular plane could be either linear, quadratic, or cubic. Tremendous amount of efforts need to be deployed to understand the unique electronic properties of such a large zoo of semi-metals and some work has already been done in this direction \cite{MANDAL2021127293,Mandal_2020,bera2021floquet,Mandal_2020higher_spin}. 

We wish to make another step in a direction that will help classify properties of the large zoo of semimetals. Specificially, we wish to borrow some ideas from statistical mechanics, where physical systems are classified according to phases and phase transitions. Phase transitions in classical and quantum systems are usually characterized by non-analytic behavior of one or more physical quantities. A sudden jump in the free energy, hereby, is classified as a first order phase transition, jumps in first or higher derivatives are classified as second order phase transitions \cite{anderson2018basic} - a classification that goes all the way back to Paul Ehrenfest \cite{jaeger1998ehrenfest}. Recently it has been pointed out that in close analogy with thermodynamic phase transitions, dynamical systems can also exhibit similar non-analytic behavior at a critical time, rather than a critical temperature or other parameter. Such behavior has been coined a dynamical phase transition \cite{heyl2013dynamical,heyl2018,PhysRevB.93.085416,PhysRevB.93.144306,flaschner2018observation} and crucially differs from the statistical mechanics case in that it can be extracted from the time evolution operator rather than a statistical ensemble density matrix. From the mathematical point of view, this similarity can be traced to the correspondence between the time evolution operator and the Gibbs ensemble density matrix through a Wick rotation. Thus, classical thermodynamic phase transitions and the dynamical phase transitions seem to be tightly related. However, experimental support to this assertion had to await recent advances in experimental techniques that opened the door to a lot of research activities in this field \cite{PhysRevB.93.144306,polkovnikov2011colloquium,flaschner2018observation,PhysRevB.96.180303,PhysRevB.96.180304,PhysRevLett.119.080501,PhysRevB.95.075143,zvyagin2016dynamical}, including experimental observations of such transitions \cite{PhysRevLett.119.080501,flaschner2018observation} as well as in the identification of exactly solvable models \cite{zvyagin2016dynamical}.

In our present work we will show that similar non analytic behavior is also present in the tunneling coefficients of anisotropic Multi-Weyl semimetals. The usual thermodynamic limits with large system size $N$, here, is replaced by the large distance limit in the transmission region beyond a localized scattering potential. 
Our paper is organized as follows. In section \ref{sec:model}, we give a brief mathematical formulation of our model Hamiltonian with linear energy dispersion along the $z$-axis and $n$-th order dispersion in the $x$-$y$ plane. In section \ref{sec:tunneling_phase_diagram} we study the tunneling problem through a square barrier and in particular investigate how the number of propagating solutions depend of the incident angle $\theta$ and an anisotropy parameter $\lambda$. We then investigate a formal connection between tunneling phase diagrams and phase transitions. In section \ref{sec:tunnelingcoeff} we consider tunneling in the case of a quadratic energy dispersion in the $x$-$y$ plane for both isotropic and anisotropic realizations and confirm non-analytical behavior near phase boundaries and classify the transitions as second order phase transitions. Finally, in section \ref{sec:conclusion} we present our conclusion and summarize our main findings.

%%%%%%%%%%%%%%%%%%%%%%%%%%%%%%
\label{sec:intro}

\section{Model}
\label{sec:model}
In  a seminal paper Fang {\it et al.} \cite{PhysRevLett108266802}
have made a prediction of Multi-Weyl semimetals that are stabilised via point group symmetries. Motivated by this work we consider the general class of Hamiltonians that are of Weyl-type and have a form
\begin{equation}
	H=\begin{pmatrix}
	v_zk_z&v_x(k_+^n+k_-^n)+v_y(k_+^n-k_-^n)\\
	v_x(k_-^n+k_+^n)+v_y(k_-^n-k_+^n)&-v_zk_z
	\end{pmatrix},
	\label{eq:Hamiltoniann}
\end{equation}
with a $n$-th order dispersion in the $x$-$y$ plane. Hereby, ${\hat{k}}_{\pm}=\hat{k}_x \pm i \hat{k}_y$ and $\hat{k}_{x,y,z}$ are momentum operators with independent real coefficients $v_{x,y,z}$. We should note that the cases $n=1,2,3$ are well justified via point group symmetry arguments \cite{PhysRevLett108266802}. For case $n>3$ we are, however, aware of specific realizations with $v_x=v_y$ and $v_z=0$ in the case of multilayer graphene \cite{10.1143/PTPS.176.227}. Nevertheless, we will first focus on the general case before restricting our attention to the case $n=2$ later.
We should also note that the model for $n=3$ is similar to the cubic Dirac semimetal that was recently predicted based on first principles calculations in \cite{PhysRevX.7.021019} and where some of its tunneling properties were discussed in \cite{BOUHLAL2021168563}. We also note that recently tunneling properties for a similar class of models - just with $v_x=v_y$ - have been studied in \cite{MANDAL2021127293}.

Therefore, for the purposes of this article we will focus on the new physics that is related to an anisotropy parameter $\lambda=v_x/v_y$. Hence, for simplicity we will consider the case of a material that is sufficiently thin that the linearly dispersed $z$-direction is frozen out and we can safely set $k_z=0$ (see the appendix in \cite{BOUHLAL2021168563} for details). The Hamiltonian can then be expressed in unitless form only in terms of the anisotropy parameter $\lambda$ as
\begin{equation}
	H_{\mathrm{flat}}=\begin{pmatrix}
	0&\frac{\lambda+1}{2}k_+^n+\frac{\lambda-1}{2}k_-^n\\
	\frac{\lambda+1}{2}k_-^n+\frac{\lambda-1}{2}k_+^n&0
	\end{pmatrix}
	\label{eq:Hamiltonian}
\end{equation}

%%%%%%%%%%%%%%%%%%%%%%%%%%%%%%%%%%%%%%%%%%%%%%%%%%
\section{Tunneling phase diagrams}
\label{sec:tunneling_phase_diagram}
\begin{figure}[htb]
	\centering
	\includegraphics[width=0.7\linewidth]{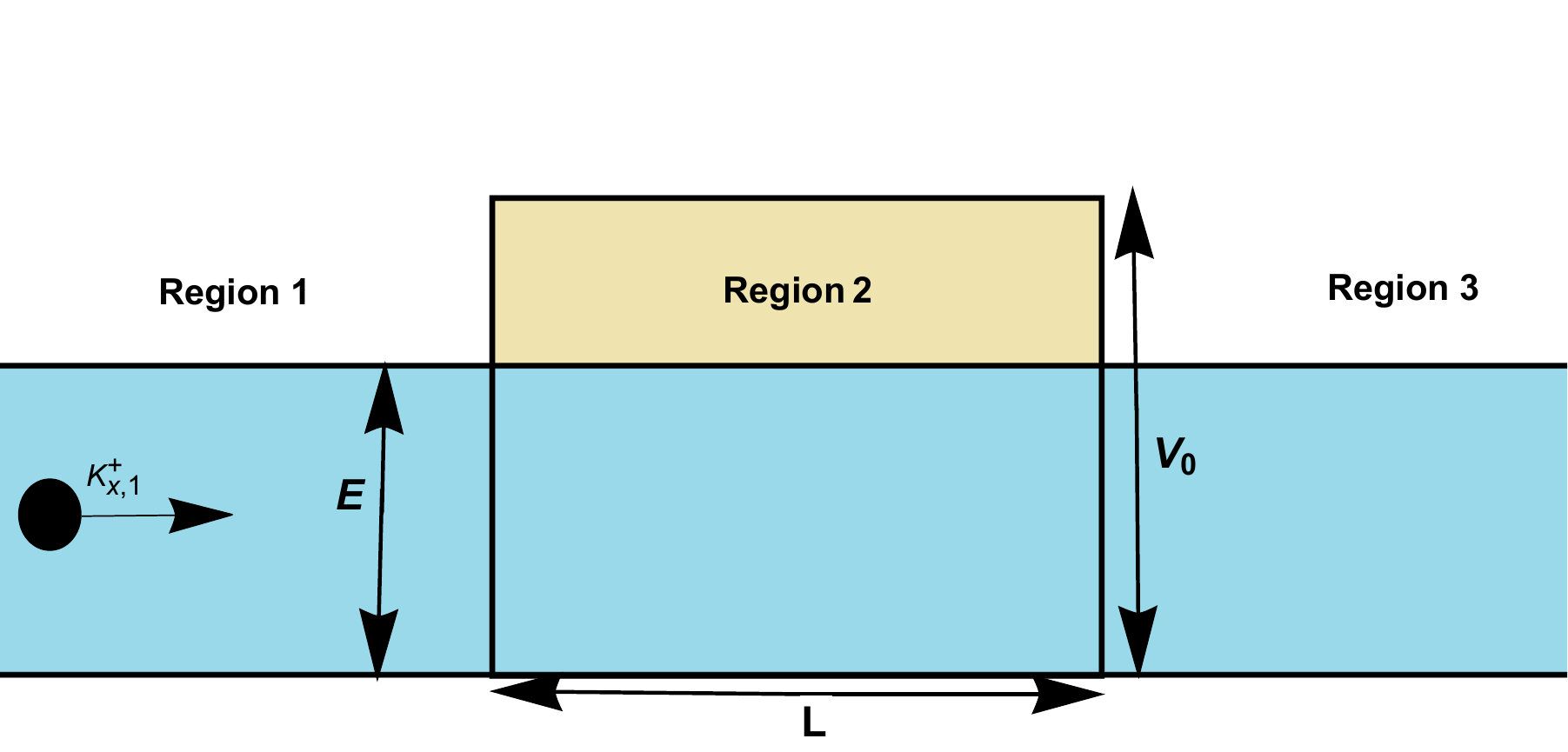}
	\caption{(color online) Shown is a cartoon of the tunneling problem with an incoming particle with momentum $k_{x,1}^+$, that will have an energy $E$. The particle is subjected to a square barrier of height $V_0$ and width $L$. The problem can be broken into free space regions 1 and 3 and a tunneling region 2.}
	\label{fig:barrierprobpic}
\end{figure}

%%%%%%%%%%%%%%%%%%%%%%%%%%%%%%%%%%%%%%%%%%%%%%%%%%
We may now consider the system subjected to square-shaped barrier region - this can be a potential or similar - in the $x$-direction. The setup is given by a function
\begin{equation}\label{eV0}
	V(x)=
	\left\{%
	\begin{array}{ll}
		 V_0, & -\frac{L}{2}<x<\frac{L}{2}\\
         0,  & \mbox{elsewhere} \\
	\end{array}%
	\right.
\end{equation} as presented in Fig. \ref{fig:barrierprobpic} below

In this work we want to study the tunneling properties of electrons in this system on general grounds. In its simplest iteration $V(x)$ is just an ordinary potential barrier. But let us for now remain more general before we restrict ourselves to this case. 
Let us assume an incoming wave with wavevector $\vect k=k(\cos\theta,\sin\theta)$. In this case for our model in free space we find an energy 
\begin{equation}
	E=\pm\frac{k^n} {\sqrt{2}} \sqrt{1+\lambda ^2+\left(\lambda ^2-1\right) \cos (2 n\theta  )}.
	\label{eq:energy_eq}
\end{equation}
A plot of the positive energy solution for $n=2,3$ is shown in Fig. \ref{fig:quadraticenergycontourplot} 
below, which displays $2n$- fold rotational symmetry.

\begin{figure}[H]
	\centering
	\subfloat[][Contour plot for the quadratic case $n=2$ case.]{\includegraphics[width=0.48\linewidth]{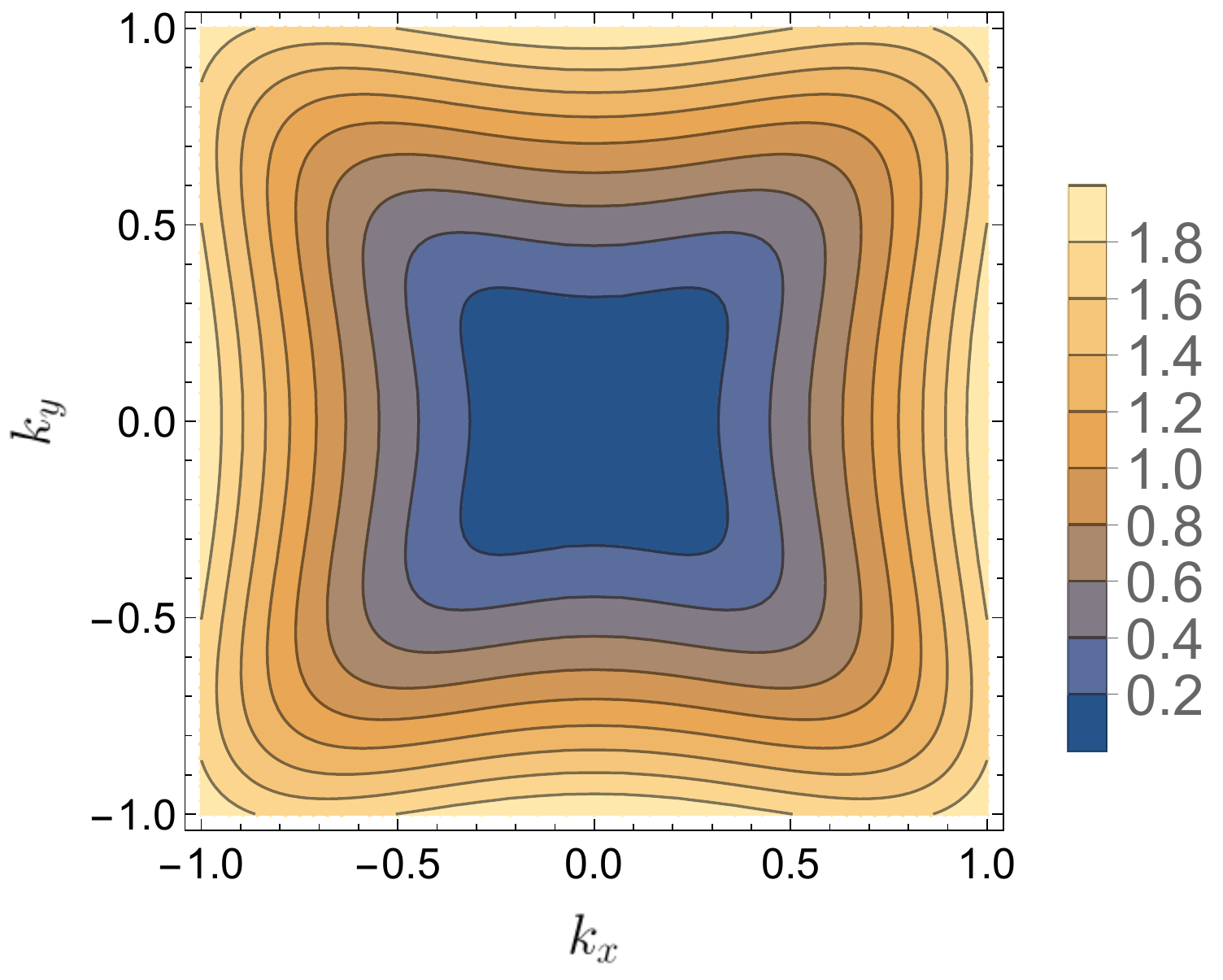}}\hspace{0.03\linewidth}\subfloat[][Contour plot for the cubic $n=3$ case.]{\includegraphics[width=0.48\linewidth]{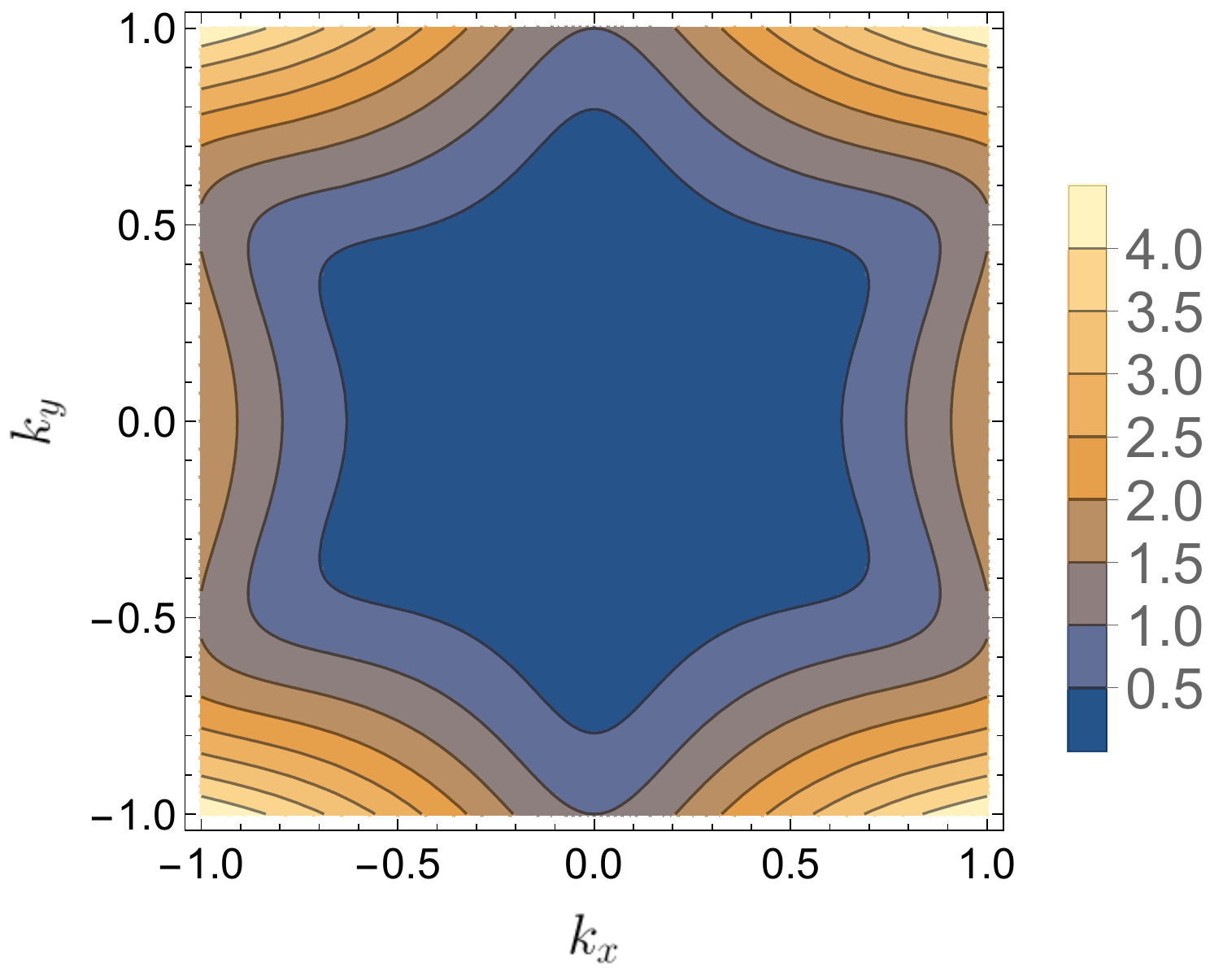}}\\
		\subfloat[][Zero $k_y$ slice for the $n=2$ case.]{\includegraphics[width=0.5\linewidth]{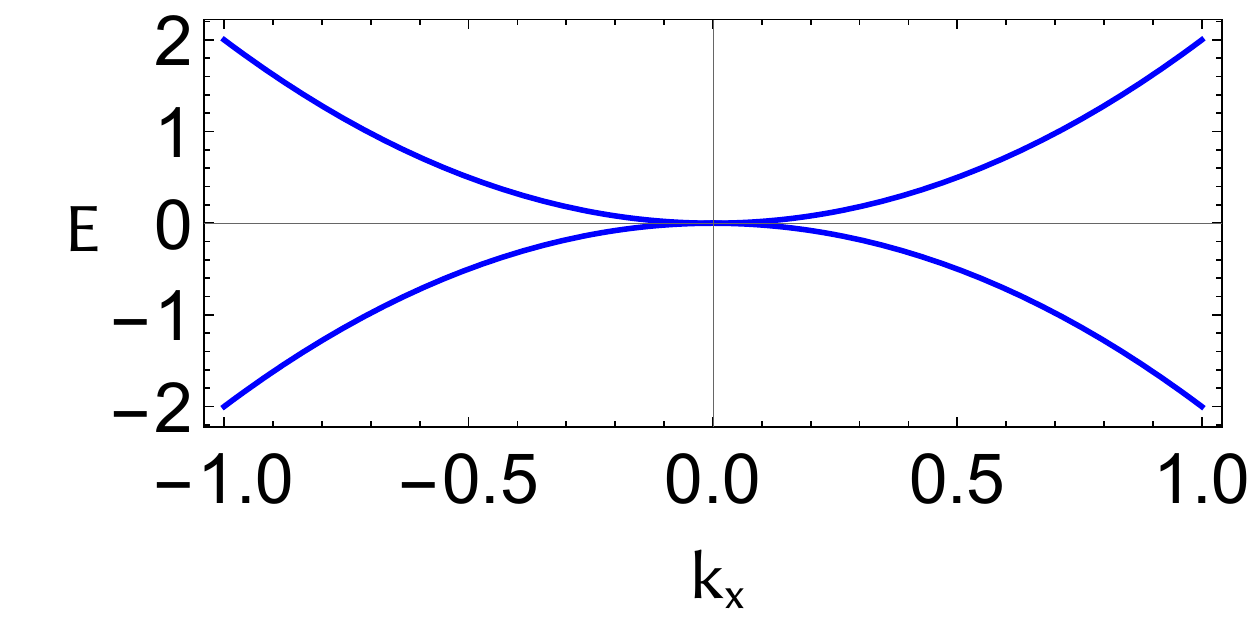}}\subfloat[][Zero $k_y$ slice for the $n=3$ case.]{\includegraphics[width=0.5\linewidth]{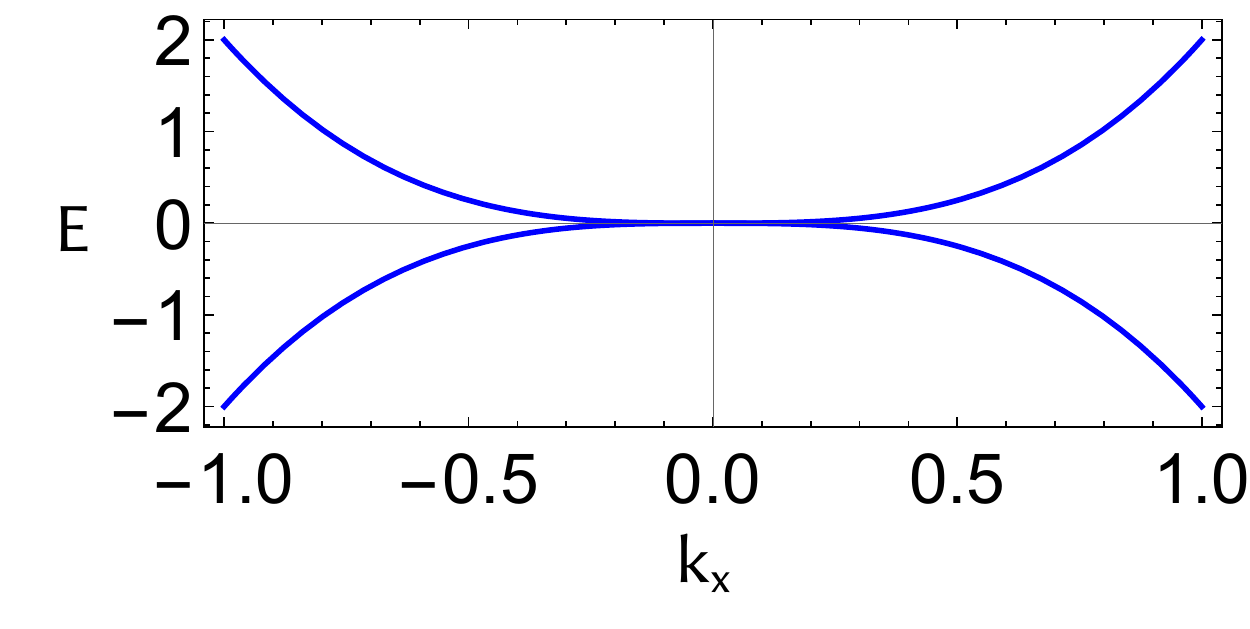}}
	\caption{(color online) Shown in the top of the figure are contour plots of the positive energy solution for $\lambda=2$ of the Hamiltonian Eq. \eqref{eq:Hamiltoniann} for different order dispersions. At the bottom we show a corresponding slice at $k_y=0$, which includes both positive and negative energy solutions.}
	\label{fig:quadraticenergycontourplot}
\end{figure}

We will find that it is the anisotropic character of the dispersion that can lead to some interesting tunneling phenomena. To gain a better understanding of this the eigenvalue problem may now be recast as a non-linear eigenvalue problem for $k_x$ at a fixed energy $E$ - assuming that we want to consider a tunneling problem for a barrier perpendicular to the $x$-direction. For the example of $n=2$ one finds the non-linear eigenvalue problem
\begin{equation}
(\lambda k_x^2\sigma_x	-2k_xk_y\sigma_y-\lambda k_y^2\sigma_x-E\mathbb{1}_2)\psi=0.
\end{equation}
Solving such a non-linear generalized eigenvalue equations for $k_x$ at a fixed energy allows us to determine what modes $k_x$ and corresponding eigenvectors can enter an ansatz for the tunneling problem in the different regions. That is, the solutions for a zero potential allow us to determine what modes can in principle be generated by the interaction with a barrier - whether it is a potential barrier, a region with magnetic field or otherwise. Of particular interest are modes with real valued $k_x$ because those survive the asymptotic limit $x\to \infty$, that is they could be measured by a distant detector since they don't decay exponentially (exponentially growing modes are not allowed for $x\to\pm \infty$ because they are unphysical). Below in Fig. \ref{fig:quadraticdiagram}
 we show a diagram that displays how many real valued solutions there are for $k_x$ at an energy given by \eqref{eq:energy_eq} for various values of $n$.
 \begin{figure}[htbp]
 	\centering
 	\subfloat[][Tunneling phase diagram for the $n=2$ case.]{\includegraphics[width=0.31\linewidth]{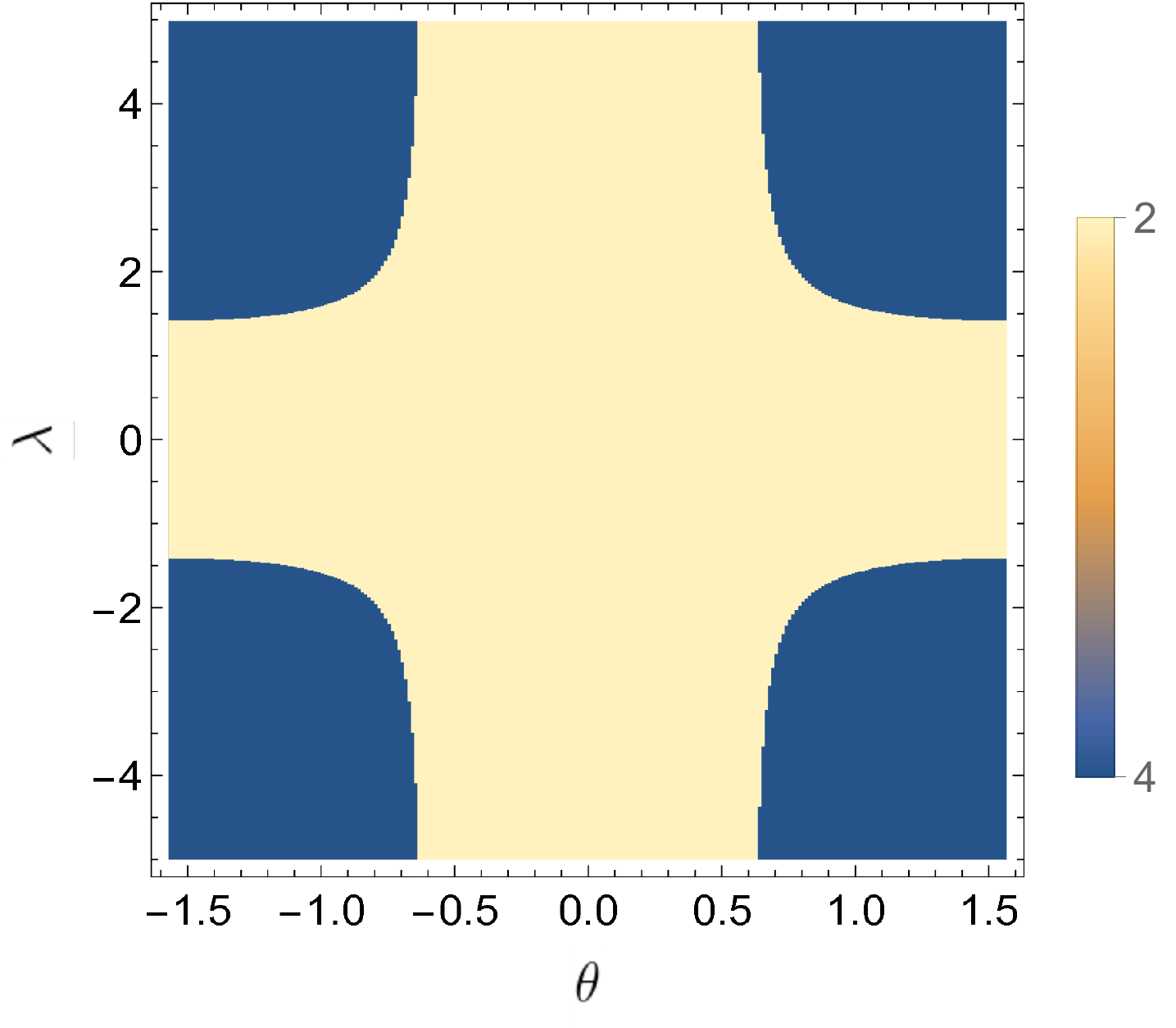}\label{fig:quadraticdiagrama}}\hspace{2cm}\subfloat[][Tunneling phase diagram for the $n=3$ case.]{\includegraphics[width=0.31\linewidth]{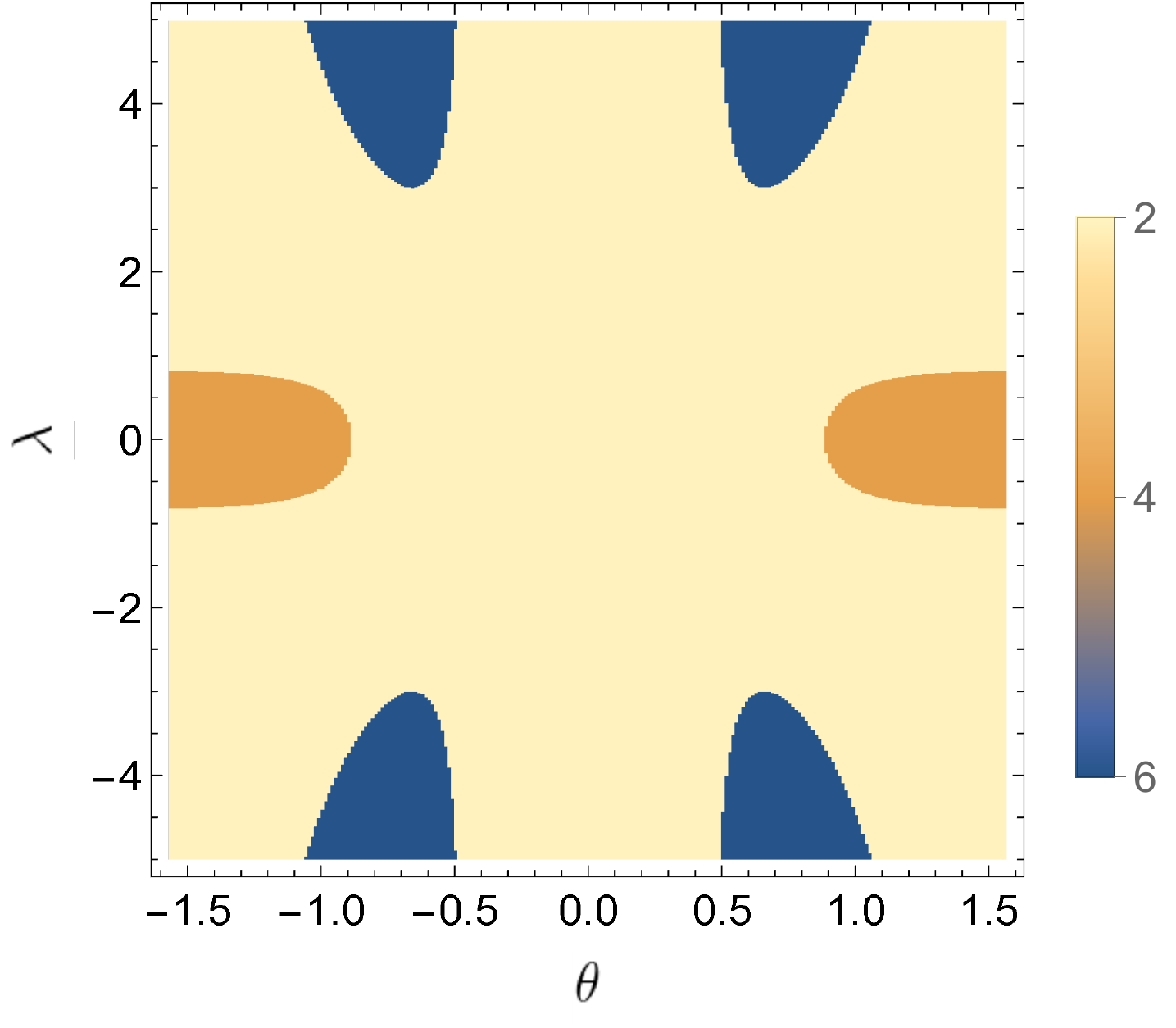}}\\
 	\subfloat[][Tunneling phase diagram for the $n=4$ case.]{\includegraphics[width=0.31\linewidth]{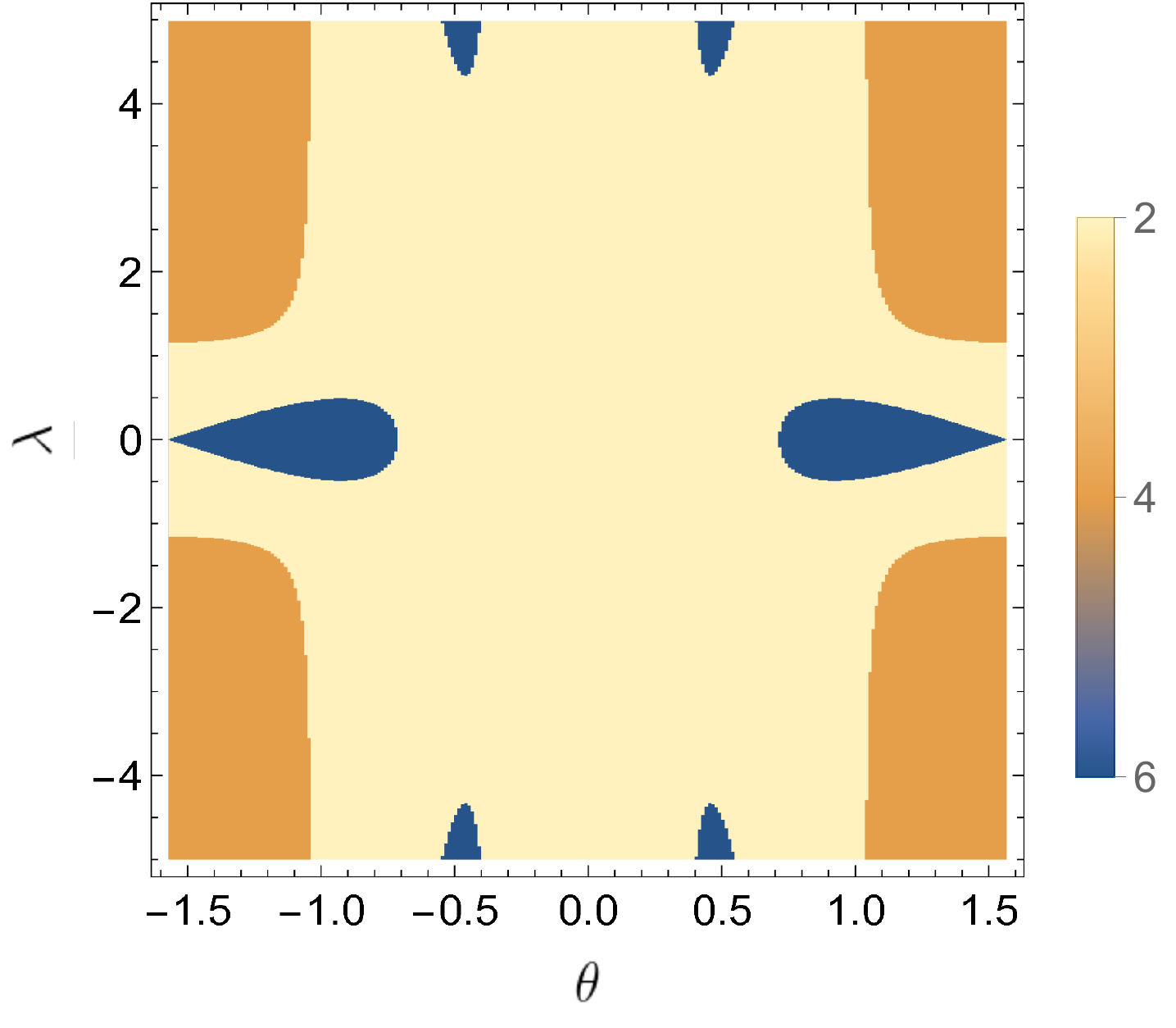}}\hspace{2cm}\subfloat[][Tunneling phase diagram for the $n=5$ case.]{\includegraphics[width=0.31\linewidth]{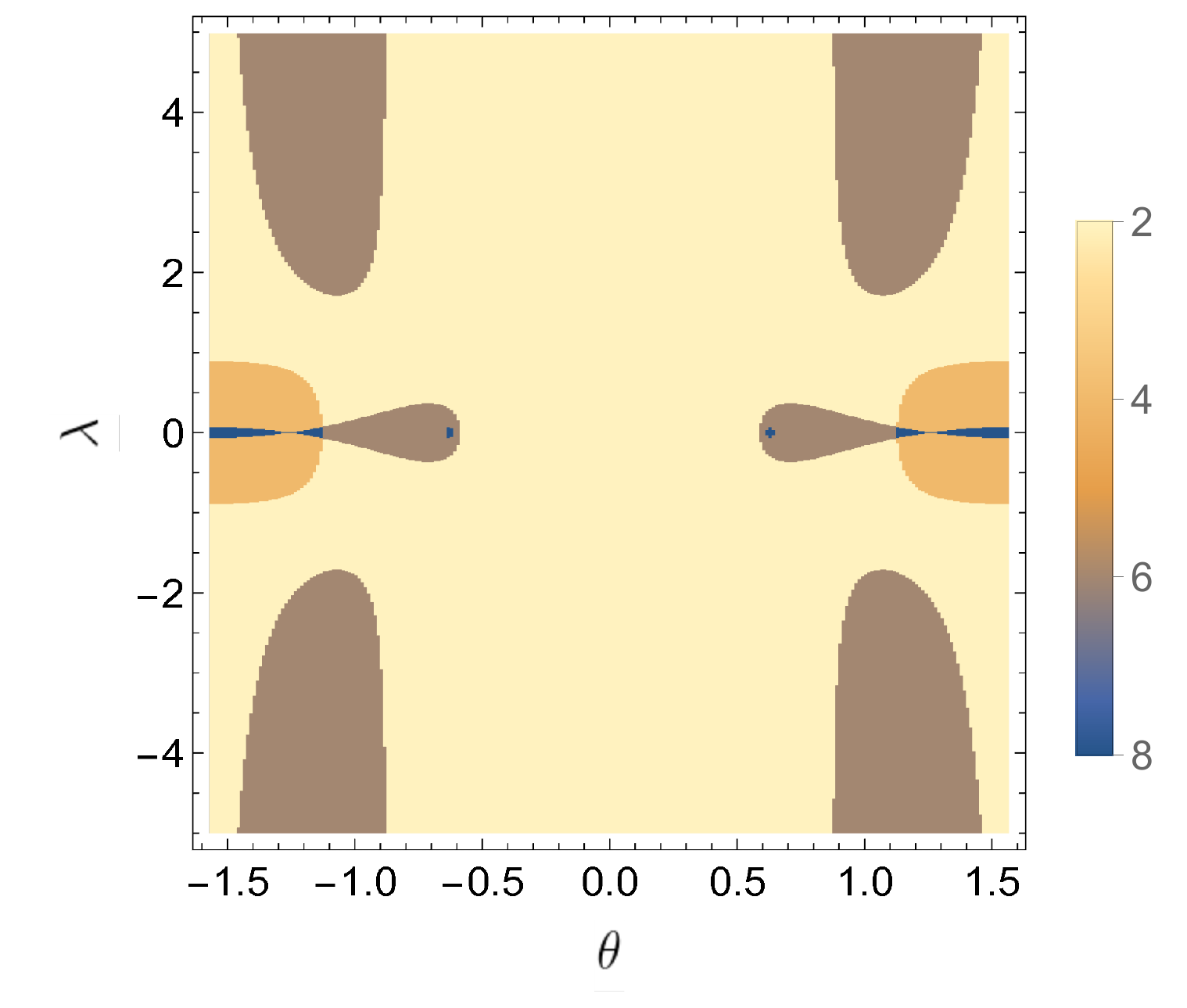}}
 	\caption{(color online) Shown is a tunneling phase diagram for the Hamiltonian \eqref{eq:Hamiltoniann} that shows how many real solutions for $k_x$ exist for certain parameter combinations of incident angle $\theta$ and anisotropy parameter $\lambda$.}
 	\label{fig:quadraticdiagram}
 \end{figure}

We find that there is different ranges of anisotropy parameters $\lambda$ and incident angle $\theta$ that allow for a different number of real valued solutions. We interpret the transitions between different regions as phase boundaries in a phase diagram of a distant detector. This suggests an interesting structure for tunneling processes, which we will interpret later. We should stress that this diagram - regardless of the specific type of barrier, whether it is a potential or something else - tells us how many modes can appear in an asymptotic limit after scattering. It is therefore a general feature of specific Multi-Weyl semimetals. One should also note the crucial dependence on the anisotopy parameter - the case of an isotropic system $\lambda=\pm 1$ is trivial with only two allowed modes regardless of incident angle.

Before we dig any deeper in our interpretation of these diagrams, let us make more clear why we use the notion of the phase diagram of a distant detector. This can be made slightly more precise by making use of an analogy to dynamical phase transitions. It has been argued in \cite{Heyl_2018}
that expressions like
\begin{equation}
P_\alpha=\left|\langle\psi_n|U(t)\psi\rangle\right|^2
\label{eq:amplitude}
\end{equation}
bear a formal resemblance to partition functions (skipping all technical details and subtleties like time ordered exponentials etc. The time evolution operator $U(t)=e^{iHt}$ is related to the Gibbs ensemble density matrix $\rho(\beta)=e^{-\beta H}$ via a Wick rotation). Non-analytic behavior that can be found in the $P_\alpha$ is at the heart of the analogy that was used to coin the term dynamical phase transition. Similarly, in the case of a tunneling problem for a distant detector one is interested in terms of the same form as \eqref{eq:amplitude} - just for the case $t\to \infty$ - rather than finite $t$. It therefore seems reasonable to expect similar non-analytic behavior, which can be related to a phase transition. In the coming sections we will see that indeed we will find non-analytic properties of tunneling coefficients as suggested in our diagrams. We argue that, however, in our case a long distance limit (distance from the barrier) takes the place of a thermodynamic large $N$ limit. That this is a useful analogy to make can be seen by the following statement: The number of real valued modes as a means to characterize the system is only sensible in this long distance limit. Indeed, outside of this limit the boundaries in the diagram \ref{fig:quadraticdiagram} are meaningless - much like the phase boundaries in thermal transitions without the thermodynamic limit.

\section{Tunneling in the case of \texorpdfstring{$n=2$}{n=2}}
\label{sec:tunnelingcoeff}
To actually understand our phase diagrams better, we will now consider the simplest case $n=2$ with a step-wise potential barrier such that the Hamiltonian for the different regions in Fig. \ref{fig:barrierprobpic} is given as
 \begin{equation}
  \label{eq:Hamiltonianf}
  H_j(k_x,k_y) = \begin{pmatrix}
	V_j&\lambda \hat{k}_x^2-\lambda \hat{k}_y^2+2 i \hat{k}_x \hat{k}_y\\
	\lambda \hat{k}_x^2-\lambda \hat{k}_y^2-2 i \hat{k}_x \hat{k}_y&V_j
	\end{pmatrix},
\end{equation}
where $V_j=\delta_{j,2}V$.
In order to solve the eigenvalue problem we can separate variables and write the eigenspinors as
plane waves in the $y$-direction. This is due to the fact that $[H_j,k_y]=0$ requires the conservation
of momentum along the $y$-direction, then we can write $\Psi(x,y)= e^{ik_y y} \psi(x)$. 
The associated eigenspinors are solutions of
the  non-linear eigenvalue equation
  \begin{equation} \label{eqphi}
 \left[\lambda\left(k_x^2-k_y^2\right) \sigma_x-2 k_x k_y \sigma_y+\left(V_j-E_j\right) \mathbb{I}_{2} \right]\psi=0.
 \end{equation}
 
 The eigenvalues in the regions 1 and 3 (compare Fig. \ref{fig:barrierprobpic}) are found to be
 \begin{eqnarray} \label{k1}
&&k_{x,1}^+= - k_{x,1}^- =k \cos\theta\\\label{k2}
&&k_{x,2}^+ = - k_{x,2}^-= -i \frac{k}{\sqrt{2} \lambda} \sqrt{f(\theta,\lambda)},
\end{eqnarray}
where we have set
\begin{equation}
    f\left(\theta,\lambda\right)=\left(3 \lambda ^2-4\right) \cos (2 \theta )-\lambda ^2+4.
\end{equation}
The number of real modes -i.e. modes that can be measured by a distant detector - therefore solely depends on the sign of $f(\theta,\lambda)$.

Asymptotically the tunneling problem for the different colored regions in Fig. \ref{fig:quadraticdiagrama}  can be interpreted as in the Fig. \ref{fig:tunneling_scheme} below:

\begin{figure}[htb]
	\centering
	\subfloat[][Figure corresponding to the brown colored region with only one propagating reflected/transmitted mode.]{\includegraphics[width=0.45\linewidth]{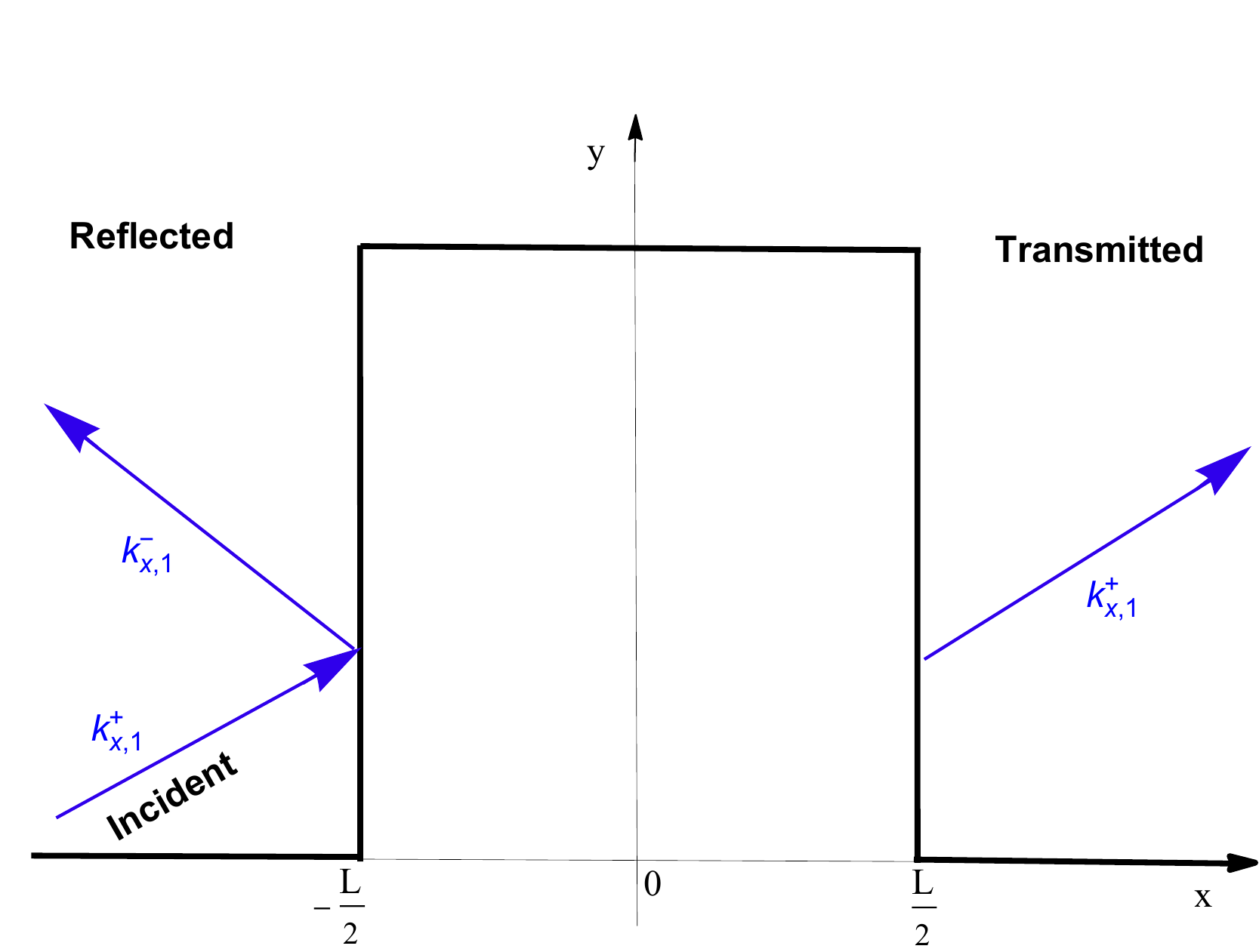}\label{fig:tunneling_schemea}}\hspace{0.09\linewidth}\subfloat[][Figure corresponding to the blue colored region with two propagating reflected/transmitted modes.]{\includegraphics[width=0.45\linewidth]{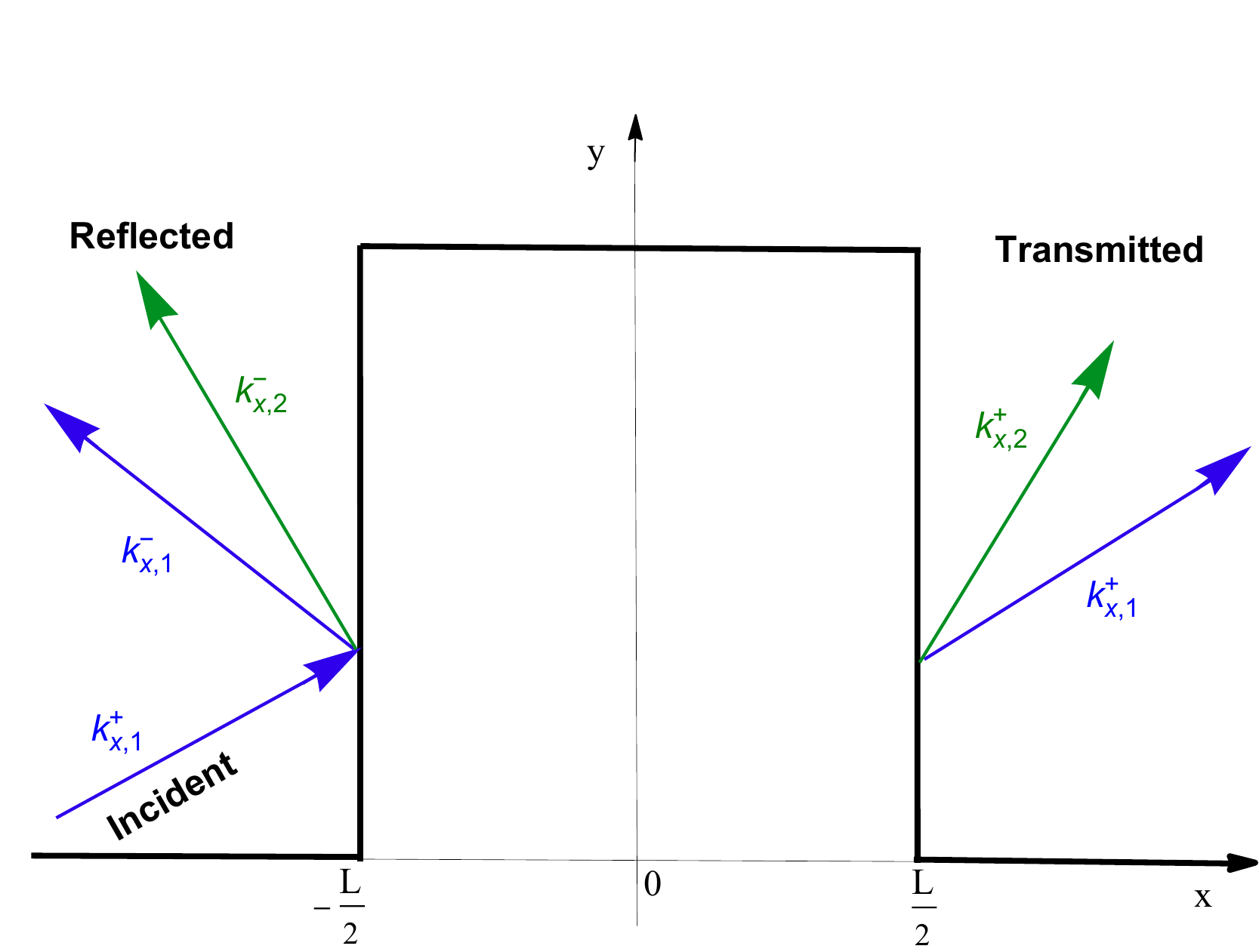}\label{fig:tunneling_schemeb}}
	\caption{(color online) Shown is a schematic diagram of incoming and outgoing waves near a potential barrier for the quadratic semimetal case. Signified by arrows we have only included wave components that survive in the asymptotic limit $x\to\pm\infty$. While not perspectively, accurate we have signified waves with different propagation directions as having different angles.  Both sides correspond to different colored regions of the phase diagram \ref{fig:quadraticdiagrama}.}
	\label{fig:tunneling_scheme}
\end{figure}

We find that in blue colored regions of Fig. \ref{fig:quadraticdiagrama} an incoming wave vector $\vect k$ can scatter into two different transmitted modes with wavevectors $(k_{x,1}^+,k_y)$ and $(k_{x,2}^+,k_y)$ (see Fig. \ref{fig:tunneling_schemeb}). That is we have transmitted waves at two different angles after a barrier. Similarly, in the brown colored regions there is only one mode that survives in the long distance limit (see Fig. \ref{fig:tunneling_schemea}). Therefore, the phase barriers in Fig. \ref{fig:quadraticdiagrama} can be interpreted as critical incident angles (at fixed incoming wavevector amplitude $k$) for which it becomes possible that there is two transmitted modes.

Similar to regions 1 and 3 (see Fig. \ref{fig:barrierprobpic}) for completeness we mention that in the region 2 we find solutions
\begin{eqnarray}\label{q1}
&&q_{x,1}^+= -  q_{x,1}^-=q \cos\phi \\\label{q2}
 &&q_{x,2}^+ = - q_{x,2}^-= -i \frac{q}{\sqrt{2} \lambda} \sqrt{f(\phi,\lambda)},
\end{eqnarray}
where $q=\left(\frac{2 \left(V_0-E \right)^2}{\left(\lambda^2-1\right) \cos\left(4 \phi\right)+\lambda^2+1} \right)^{\frac{1}{4}}$ and energy $E$ given as in Eq. \eqref{eq:energy_eq}. Furthermore, one may recall that for an incoming wave in region 1, $k_y=k\sin\theta$ and per definition $q_y=q \sin\phi$. From momentum conservation in $y$-direction we then have $\phi=\sin^{-1}(\frac{k}{q}\sin\theta)$.
 
We may now analyze tunneling properties in the system.
\subsection{Isotropic case \texorpdfstring{$\lambda=1$}{L=1}}
First recall that Fig. \ref{fig:quadraticdiagram} shows that there are two parameter regions that have to be treated separately in an analytical discussion. The blue region in the diagram has been found to host four real modes, while the light brown region hosts only two real modes and two complex modes. For simplicity and to gain some easy analytical insights let us first consider the light brown region, which only has one transmitted mode that survives the asymptotic limit $x\to \infty$. More specifically we will restrict our discussion in this section to the isotropically dispersed case $\lambda=1$. This is because most analytic progress can be made for this case and the discussion is not obfuscated by mathematical complications. The more general case for a simple discussion will only be treated numerically.
In the case $\lambda=1$, we find that the wavefunctions $\Phi_{i}(x)$ in region $i$ are given as
\begin{eqnarray}
 \label{eq:phi1}
&&\Phi_{1}(x)=  e^{ i k_{x,1}^+ x}\begin{pmatrix}
\chi_{k,1}^{+} \\ 1 
\end{pmatrix} 
+ r_1 e^{ i k_{x,1}^- x}\begin{pmatrix}\chi_{k,1}^{-} \\ 1 
\end{pmatrix} + r_2 e^{ i k_{x,2}^+ x}\begin{pmatrix}
\chi_{k,2}^{+} \\ 1 
\end{pmatrix},
\end{eqnarray}
\begin{eqnarray}  \label{eq:phi2}
\Phi_{2}(x)=b_1 e^{ i q_{x,1}^+ x}\begin{pmatrix}
\chi_{q,1}^{+} \\ -1 
\end{pmatrix} + b_2 e^{ i q_{x,1}^- x}\begin{pmatrix}
\chi_{q,1}^{-} \\ -1 
\end{pmatrix} + b_3 e^{ i q_{x,2}^+ x}\begin{pmatrix}
\chi_{q,2}^{+} \\ -1 
\end{pmatrix}+ b_4 e^{ i q_{x,2}^- x}\begin{pmatrix}
\chi_{q,2}^{-} \\ -1 
\end{pmatrix}
\end{eqnarray}
\begin{eqnarray}
\label{eq:phi3} 
\Phi_{3}(x) =t_1 e^{ i k_{x,1}^+ x}\begin{pmatrix}
	\chi_{k,1}^{+} \\ 1 
	\end{pmatrix} + t_2 e^{ i k_{x,2}^- x}\begin{pmatrix}
	\chi_{k,2}^{-} \\ 1 
	\end{pmatrix},
\end{eqnarray}
where we have used the short-hand notation
\begin{equation}
\chi_{k,m}^{\pm}=\frac{ k_{x,m}^{\pm} + i k \sin\theta}{  k_{x,m}^{\pm} - i k \sin\theta}
\end{equation} 
and $ m=1,2$ is labeling different wave vectors. Hereby, $r_i$, $b_i$ and $t_i$ are reflection, barrier and transmission coefficients that have to be determined by the boundary conditions at the barrier. 

We note that from all the possible momentum solutions shown in Eqs. \eqref{k1} and \eqref{q1} we chose the appropriate ones that are physically allowed to contribute as follows. For region 1 ($x<-{L}/{2}$), we need to have finite $\Phi_{1}$ as $x \to -\infty$ and we find that $k_{x,1}^\pm$ is a propagating wave and therefore is allowed. Next we see that $k_{x,2}^{-}$ corresponds to a solution grows exponentially as $x\to- \infty$ and so it is not physically allowed.  In region 2 ($-\frac{L}{2}<x<\frac{L}{2}$) there is no physical restrictions and hence all possible plane wave contributions can be included. Lastly, for region 3 ($x>\frac{L}{2}$), where only propagation away from the barrier is allowed and terms need to be finite for $x\to \infty$, we find that $k_{x,1}^+$ and $k_{x,2}^-$ fulfill this condition.

To determine the coefficients $r_i$, $b_i$ and $t_i$, we use the boundary conditions at the interfaces $x=\pm L/2$.  The continuity of the spinor wavefunctions and their first derivatives  at each junction interface read as
\begin{eqnarray}
&& \Phi_1\left(- \frac{L}{2}\right) = \Phi_2\left(- \frac{L}{2}\right), \qquad \Phi_2\left(\frac{L}{2}\right) = \Phi_3\left(\frac{L}{2}\right)\\
&&{\partial}_x\Phi_1(x)|_{x=- \frac{L}{2}}={\partial}_x\Phi_2(x)|_{x=- \frac{L}{2}},\qquad 
{\partial}_x\Phi_2(x)|_{x=\frac{L}{2}}= {\partial}_ x\Phi_3(x))_{x=\frac{L}{2}} 
\end{eqnarray}
After the different coefficients for the wavefunction are determined one may now determine the currents in the different regions. The current in $x$-direction is found as
\begin{equation}\label{curde}
J_x= i \left[ \left( \left(\frac{\partial}{\partial x}\Psi\right)^{\dagger} \sigma_x \Psi- \Psi^{\dagger} \sigma_x \frac{\partial}{\partial x}\Psi \right) - 2 \left(\frac{\partial}{\partial y}\Psi\right)^{\dagger} \sigma_y \Psi\right] .
\end{equation}
One may then use the transmitted current $J_x^{\mathrm{tra}}$ and incident current $J_x^{\mathrm{inc}}$ to define a transmission  and reflection coefficients. Here, one has to keep in mind that in the limit $x\to\pm\infty$ not all terms in the wavefunction will survive. Specifically, for the wavefunction  in region 1  the term $r_2$ will not survive the asymptotic limit because it decays rapidly away from the barrier. Similarly, for the wavefunction  in region 3   the term $t_2$ will not survive the asymptotic limit. Hence, we can easily define transmission and reflection coefficients in the asymptotic 
limit as ratios
\begin{eqnarray}
R_1=\left|\frac{J_x^{\mathrm{ref}}}{J_x^{\mathrm{inc}}}\right|^2=|r_1|^2, \qquad
T_1=\left|\frac{J_x^{\mathrm{tra}}}{J_x^{\mathrm{inc}}}\right|^2=|t_1|^2,
\end{eqnarray}
where $J_x^{\mathrm{inc}}$ is the $x$-component of the incoming current, $J_x^{\mathrm{ref}}$ the reflected current in the asymptotic limit and $J_x^{\mathrm{tra}}$ the transmitted current in the asymptotic limit.

For the case of normal incidence $\theta=0$ the problem permits a simple analytic solution
and the transmission coefficient $t_1$  is given by 
 \begin{equation}
 t_{1}=\frac{2 i k q e^{-i k L}}{(k-q) (k+q) \sinh (L q)+2 i k q \cosh (L q)},
 \label{eq:transm_norm_inc}
 \end{equation}
 which in the limit $L\to\infty$ becomes zero. This is a result that was previously noted by \cite{katsnelson2012}.

\begin{figure}[H]\centering
	\subfloat[][Transmission amplitude.]{\includegraphics[width=0.45\linewidth]{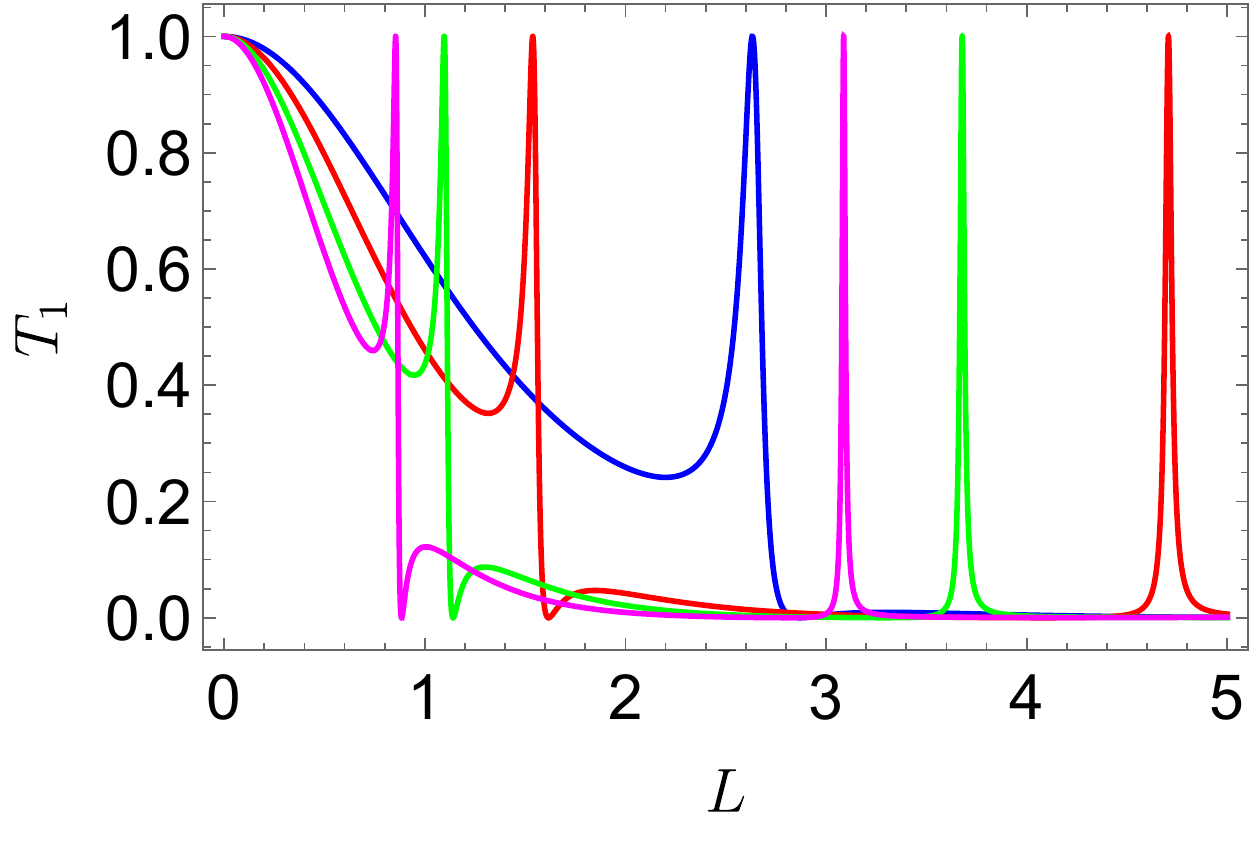}}\subfloat[][Reflection amplitude.]{\hspace{0.08\linewidth}\includegraphics[width=0.45\linewidth]{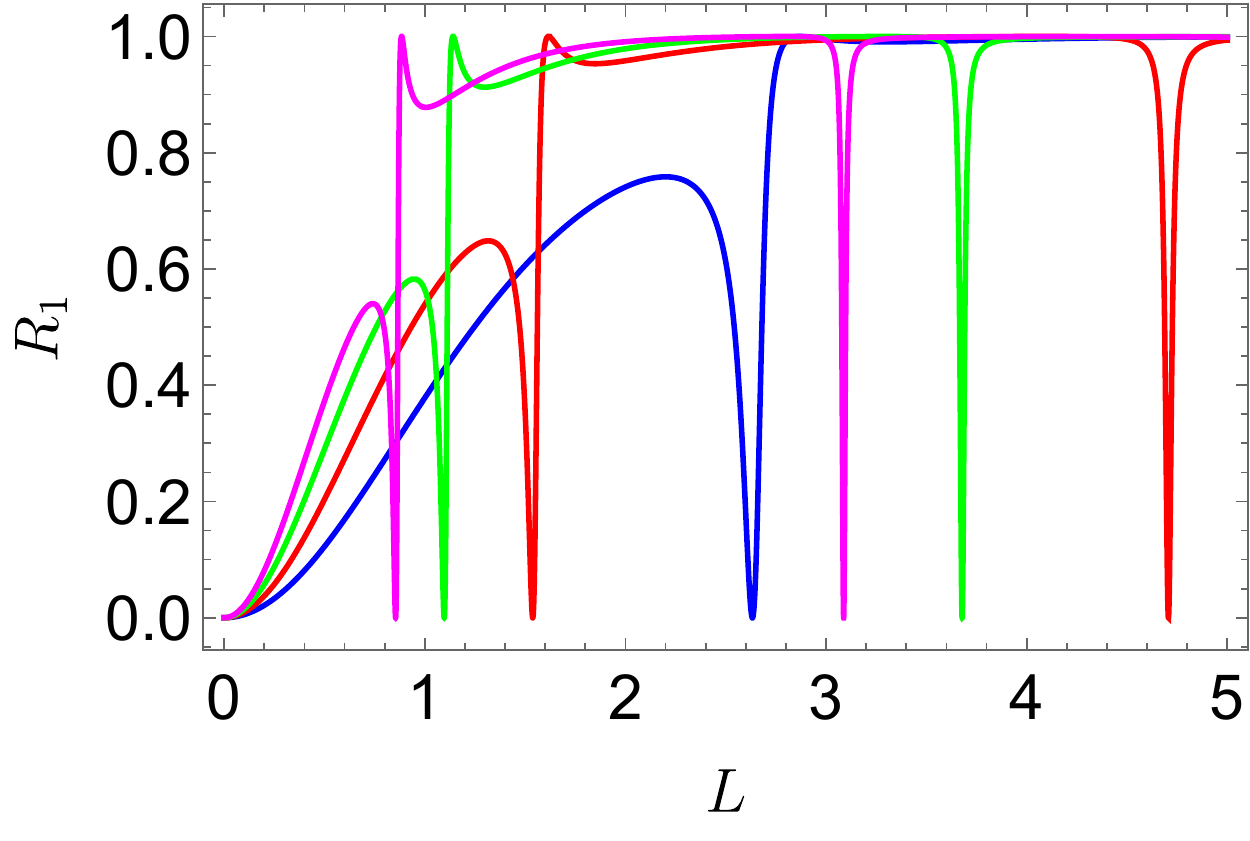}}
	\caption{(color online) Plotted is the transmission coefficient
		$T_1$ and reflection coefficient $R_1$ as a function of barrier thickness $L$.  All plots were done for constant $k=1$ and $\theta=0.1$ rad. We considered the different cases of barrier height $u_0$ with $u_0 =2$ (blue line),  $u_0 =3$ (red line ),  $u_0 =3.5$ (green line) and $u_0 =5$ (magenta line).} \label{fig:plot_length}
\end{figure}

Below we study the more general, arbitrary incident angle, but still isotropic case numerically. First we plot the transmission and reflection coefficients as a function of barrier thickness $L$, which is shown in Fig. \ref{fig:plot_length}.
We find that the the transmission coefficient $T_1$ has well-known and expected Fabry-Perot resonances. The same resonances can be seen as dips in the reflection amplitude $R_1$ because they fulfill the requirement that $T_1+R_1=1$, which follows from the conservation law for the probability current. 

\begin{figure}[H]\centering
	\includegraphics[width=0.5\linewidth]{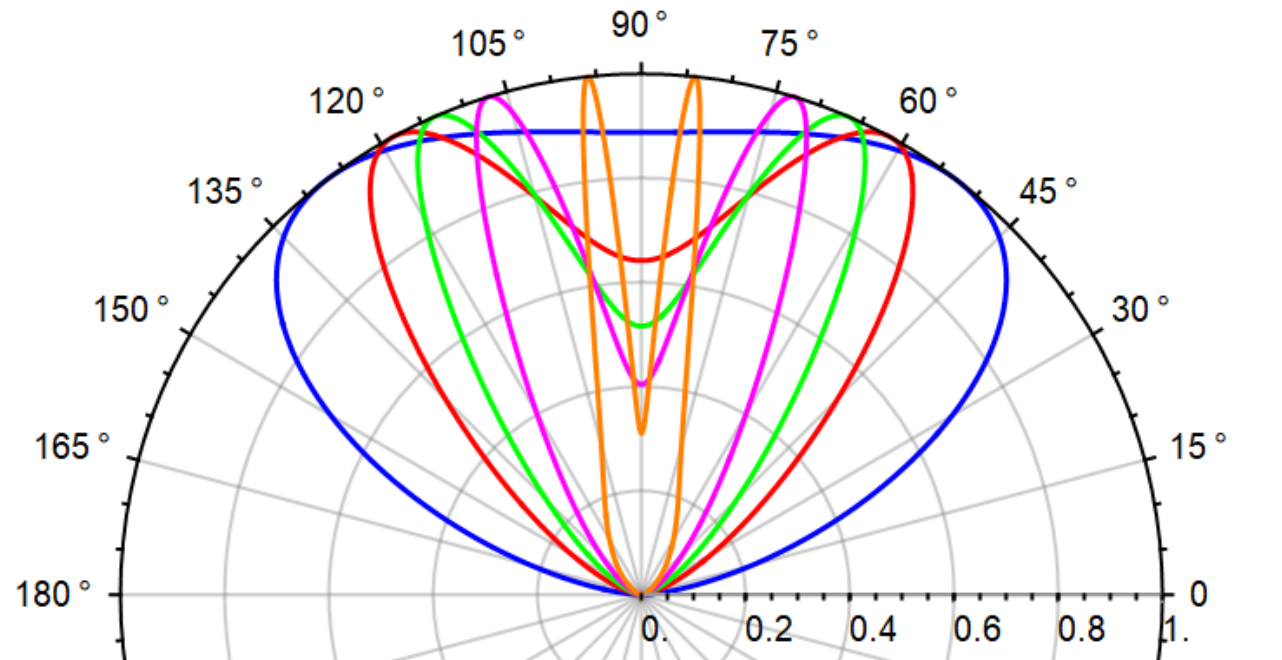}
	\caption{(color online) Shown is the transmission coefficient
		$T_1$  aa a function of incident angle incident angle $\theta$ for wave vector amplitude $k=1$, barrier height $u_0=3.5$ and for various values of  barrier thickness $L=0.2$ (blue line), $L=0.4$ (red line), $L=0.5$ (green line), $L=0.6$ (magenta line), $L=0.7$ (orange line).} \label{fig:polar}
\end{figure}

In a similar fashion we are able to see the behavior of the transmission amplitude $T_1$ as function of angle. This is shown in Fig. \ref{fig:polar}.
We find for normal incidence $\theta=0$ that there is a dip in the transmission amplitude that gets closer to zero as we increase the width $L$ of the barrier. This effect is sometimes referred to as anti Klein tunneling \cite{katsnelson2012} and is something that we can see easily from our analytical solution for normal incidence Eq. \eqref{eq:transm_norm_inc}.

\subsection{General case}
For the general anisotropic case an analytical discussion would go along the same lines as the previous section but is riddled with various technical pitfalls - for instance wavefunctions in blue colored and brown colored regions of Fig. \ref{fig:quadraticdiagram} have different expressions with different asymptotics etc. To avoid discussing these complications, which do not offer any useful insights, we rather make use of a simulation package KWANT \cite{Groth_2014} to compute tunneling amplitudes. The other reason we use this package is because it makes direct use of a description involving scattering matrices and therefore makes the relation of the phase diagram in Fig. \ref{fig:quadraticdiagram} with conventional phase transitions more lucid (see Sec. \ref{sec:tunneling_phase_diagram} for details).
More precisely, we use numerical methods to express the scattering matrix and then extract transmission/reflection amplitudes at a given energy. Our first step in numerically solving the Schr\"odinger equation is to discretize the Hamiltonian and express it in terms of a tight binding model with onsite potentials $V_{mn}$ and hopping parameters $t_{ab}$. In our particular case, the system is infinite in the $y$-direction, which allows us to use the translational invariance and treat $k_y$ as a parameter.  This means that for any fixed $k_y$, we obtain the band structure as function of $k_x$. 
More explicitly, the discretization we chose is a square lattice  tight binding model given as
\begin{equation}
    H=\sum_{n,m}\left[V_{nm}c_{n,m}^\dag c_{n,m}+\sum_{a,b=-1}^1c_{n,m}^\dag t_{a,b}c_{n+a,m+b}\right],
\end{equation}
with nearest (NN) neighbor hopping matrices
\begin{equation}
    t_{1,0}=-t_{0,1}=-\lambda\sigma_x,
\end{equation}
and next nearest neighbor (NNN) hopping matrices 
\begin{equation}
    t_{1,-1}=-t_{1,1}=\frac{1}{2} \sigma_y,
\end{equation}
 the remaining  hoppings are zero.
Hereby, $\sigma_i$ are Pauli matrices, $c_{n,m}=(c_{n,m,\uparrow},c_{n,m,\downarrow})$ is a vector of annihilation operators and $c_{n,m}^\dag$ is the corresponding vector of creation operators. The labels $\uparrow/\downarrow$ correspond to the pseudospin degree of freedom in Eq. \eqref{eq:Hamiltonian}.

%The NN hopping parameter in the x-direction is $t_x=-\lambda\sigma_x $ whereas it equals   $t_y=+\lambda\sigma_x $ in the y-direction. We complete the NNN hopping parameters by putting $t_{xx}=-\frac{1}{2} \sigma_y$ and $T_{x-x}=+\frac{1}{2} \sigma_y$ in the directions $(1,1)$ and $(1,-1)$ respectively.  
As a test of this numerical approach and our discretization in particular, we verified that the numerically computed $E(k_x)$ agrees with the analytical result $E=\sqrt{4\lambda^2 (\cos{k_x}-\cos{k_y})^2+(2 \sin{k_x}\sin{k_y})^2}$ (for $n=2$ in Eq. \eqref{eq:Hamiltonian}) in the low energy regime that we will restrict ourselves to. That this agreement is indeed good can be seen in  Fig. \ref{fig:band} below:
\begin{figure}[htb]
	\centering
	\includegraphics[width=0.55\linewidth]{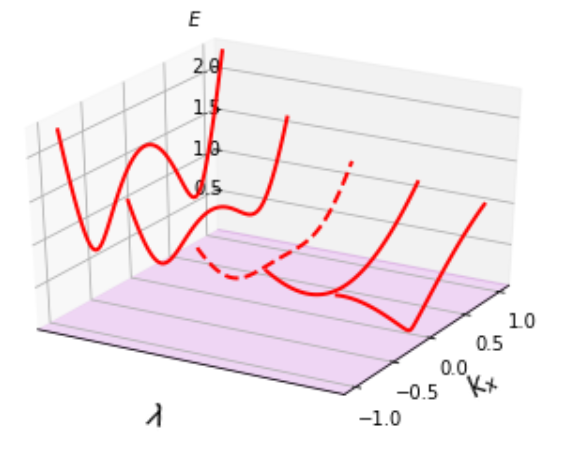}
	\caption{(color online) Different bands for different values $\lambda$ at the same value $k_y=0.65$ (in units of $1/a$ and $a$ is the lattice constant of the tight binding model we used). The dashed line is obtained for $\lambda=\sqrt{2}$. Possible scattering to two modes $\lambda>\sqrt{2}$ (left curves) and scattering to a unique mode for $\lambda<\sqrt{2}$ (right curves). }
	\label{fig:band}
\end{figure}

Our numerical results agree with our analytical prediction in Fig. \ref{fig:quadraticdiagrama} that for $\lambda<\sqrt{2}$ an incoming mode can be scattered to a unique mode with the same angle. In the other case  $\lambda>\sqrt{2}$ the transmission to a second mode is possible as it can be seen in Fig \ref{fig:band}. To understand why this happens, we need to recall that since the Hamiltonian is not isotropic, changing the incident angle will lead to a change in the energy (at a fixed wavenumber $k$) and thus, for a critical angle $\theta_c$ a change in number of allowed modes may occur. An example of how this effect manifests itself in a transmission amplitude computed at a fixed incoming wavevector $\vect k=k(\cos\theta,\sin\theta)$ can be seen in Fig. \ref{fig:Transmission} below.
\begin{figure}[htbp]
	\centering
	\includegraphics[width=1\linewidth]{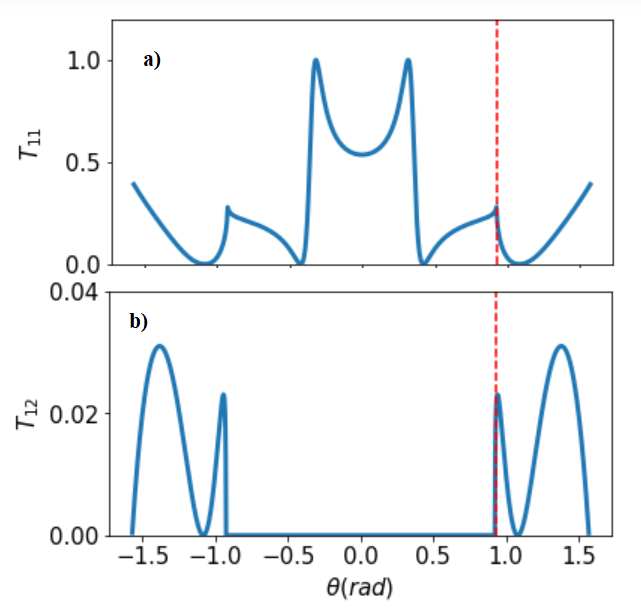}
	
	\caption{(color online) (a): The transmission $T_{11}$ from one mode to the same mode as function of the incoming particle angle. (b): the transmission $T_{12}$ from the incoming mode to another mode. The dashed line is the angle for which  a second mode opens up ($\lambda=1.67$). As parameters we used the potential $v=0.066 1/a^2$ and $L=2a$, $k=0.067/a$, where $a$ is the lattice constant of the tight binding model. For these conditions the energy $E$ is not constant (wavenumber $k$ is kept constant) but it is chosen to be smaller than the barrier potential $E<V$ for all incident angles (tunneling regime).}
	\label{fig:Transmission}
\end{figure}

We first note that the transmission to the same mode is several magnitudes larger than the transmission to the second mode. We also find that although we are in a tunneling regime, full transmission is still possible. Indeed, two maxima with unit transmission appear in Fig. \ref{fig:Transmission}{\color{blue}a}. 
Importantly, for our discussion of a tunneling phase diagram, we find that the transmission $T_{12}$ to the second mode indeed appears at the critical angle (marked via red dashed line in Fig. \ref{fig:Transmission}) that was predicted by Fig. \ref{fig:quadraticdiagrama}. Furthermore, we note the non-analytical behavior near the transition point. Markedly the transition - while non-analytic - is not a jump but rather a jump in derivative, which allows us to identify this behavior as second order transition. We call it a second order tunneling phase transition, which allows us to complete our analogy between the tunneling phase diagram Fig. \ref{fig:quadraticdiagram} and ordinary phase diagrams. Therefore, we are fully justified in using the term phase diagram to describe Fig. \ref{fig:quadraticdiagram}.

% will get closer to each other as $\lambda$ approaches  $\sqrt{2}$ and then merge to a unique maximum for $\lambda =\sqrt{2}$. 

\section{Conclusion}
\label{sec:conclusion}

In conclusion we have studied tunneling in non-isotropic Multi-Weyl semimetals and found them to have interesting tunneling phenomena such as Klein tunneling, anti-Klein tunneling and the possibility for an ordinary potential to scatter electrons into two or more transmitted modes with different directions of wave propagation. We found that this type of phenomenon occurs only after certain critical values of an anisotropy parameter and incident angle. We summarized this observation in Fig. \ref{fig:quadraticdiagram}, which we were able to identify as a kind of phase diagram - invoking an analogy to a time dependent phase transition. This analogy allowed us to coin the term tunneling phase transition because we found that - much like an ordinary phase transition - phase boundaries in Fig. \ref{fig:quadraticdiagram} coincide with non-analytical behavior in experimental observables (in our case the tunneling coefficients in Fig. \ref{fig:Transmission}), which is the hallmark of any phase transition.

Unlike, the case of more conventionally studied phase transitions the non-analytical behavior was not caused by a large number of particles $N\to \infty$ limit but rather by a long distance limit $L\to \infty$. We anticipate that these observations can serve as motivation to look for similar phenomena in otherwise unexpected places. Afterall, our work shows that the large $N$ limit is not unique in being able to cause behavior akin to phase transitions of matter. Rather, similar behavior can be expected in other places with limiting behavior.

\section*{Acknowledgments} H.B. and M.V. gratefully acknowledge the support provided by the Deanship of Research Oversight and Coordination (DROC) at King Fahd University of Petroleum \& Minerals (KFUPM) for funding this work through start up project No.SR211001.
\bibliographystyle{elsarticle-num}
\bibliography{literature}
 \end{document}